\documentclass[preprint,amsmath,amssymb,pra,aps]{revtex4-1}
\usepackage{graphicx}
\usepackage{color} 
\usepackage{graphics}
\usepackage{amsmath}
\usepackage{dcolumn}
\usepackage{amssymb}
\usepackage{bm}
\usepackage[latin1]{inputenc}

\begin{document}
\title{Gap dependent mass of photon in photonic topological insulator}
\author{Marcelo Vieira}
\email{marcelovieira@ccea.uepb.edu.br}
\affiliation{ Centro de Ci\^encias
Exatas e Sociais Aplicadas, Universidade Estadual da Para\'{\i}ba, Patos, PB, Brazil}
 \author{Sergei Sergeenkov}
\email{Electronic address: sergei@fisica.ufpb.br}
\author{Claudio Furtado\footnote{Phone: +55 83 3216 7534}}
\email{Electronic address: furtado@fisica.ufpb.br}
\affiliation{Departamento de F\'{\i}sica, Universidade Federal da Para\'{\i}ba, Caixa Postal 5008, 58051-970, Jo\~ao Pessoa, PB, Brazil}


\begin{abstract}
By using an analogy with axionic like systems, we study light propagation in periodic photonic topological insulator (PTI). The main result of this paper is an explicit expression for the PTI band structure. More specifically, it was found that for nonzero values of the topological phase difference $\gamma=\theta_2-\theta_1$ a finite gap  $\delta \propto\gamma^2$ opens in the spectrum which is equivalent to appearance of nonzero effective photon mass $m^{*}(\delta)\propto \frac{\sqrt{\delta}}{\delta +2}$. 
\end{abstract}


\maketitle

\section{Introduction}

Photonic crystals are artificial periodic structures in which electromagnetic waves (photons) can be manipulated in an analogous way to the electrons in semiconductors \cite{sciam,ho}. These structures may represent a combination of dielectric and magnetic materials \cite{jono} including more sophisticated configurations exhibiting non-trivial properties \cite{apl,prb}. In particular, photonic crystals can be used as devices for confinement and light curving \cite{bend} as well as a single frequency filter and a cavity for nanolasers \cite{fabry}. Unlike the conventional optical fibers (where the light propagates by successive total reflections leading to a significant loss of energy), in the so-called photonic fibers the light follows a practically loss-free path. 
A topological insulator (TI) is a material with rather non-trivial properties. On one hand, it is an insulator in the bulk, while on the other hand it is an electric conductor on the surface \cite{topol}. 
In the bulk of a topological insulator, the electronic band structure behaves like an ordinary band insulator, with the Fermi level being between the conduction and valence bands. On the surface of a topological insulator there are special states which allow the surface metallic conduction. The electrodynamic properties of the topological insulators closely follow the behavior predicted for the so-called axions \cite{axion}, including, for example, \textit{magneto-electric effect} which has been observed in 3D topological insulators \cite{nature}. 
The discovery of topological insulators  \cite{s1,monopolo,s2} has rekindled interest in existence of nontrivial topological phases in optical systems. 
In particular, photonic counterparts of quantum Hall edge states have been predicted \cite{s3,s4} and experimentally observed \cite{s5,s6,s7} in systems with broken time-reversal symmetry. 
  
In this paper we propose an exactly solvable  model of the photonic topological insulator (PTI) and discuss its unusual properties. In particular, an explicit
 expression for the PTI band structure (spectrum) $\omega(k)$ with TI induced gap parameter $\delta$ has been obtained, leading to the gap dependent
  effective photon mass $m^{*}(\delta)$.  The potential candidates for manifestation of the predicted in this paper phenomena include such materials as \cite{s1}  $Bi_{1-x}Sb_x$, $Bi_2Se_3$, $Bi_2Te_3$ and $Sb_2Te_3$ as well as recently studied strained thin $HgTe$ films embedded between two $Cd_{0.7}Hg_{0.3}Te$ layers \cite{nature}.

\section{The Model}

Recall that the electromagnetic response of an ordinary (nontopological) insulator is described by the Maxwell action
\begin{equation}
S_0=\int d^3xdt\left(\epsilon_0\epsilon{\bf E}^2-\frac{1}{\mu_0\mu}{\bf B}^2\right),
\end{equation}
 where {\bf E} and {\bf B} are electric and magnetic fields, leading to conventional macroscopic fields describing electric polarization and magnetization
\begin{equation}
{\bf D}=\frac{\delta S_0}{\delta {\bf E}}=\epsilon_0\epsilon(\textbf{r}){\bf E},
\end{equation}

\begin{equation}
 {\bf H}=\frac{\delta S_0}{\delta {\bf B}}=\frac{1}{\mu_0\mu(\textbf{r})}{\bf B},
 \end{equation}
Here,  {\bf D} is the electric field displacement and {\bf H} is the magnetic field intensity.

What would be the electromagnetic response of a topological insulator (TI)?  The novel property of the TI is that it has a conductive surface. It means that for an incident electric field, a surface current will arise inducing a magnetic field in the material, that is magnetizing it. In other words, in the TI an electric field induces a magnetization. Likewise, the magnetic field induces a dielectric polarization in the TI. These two phenomena result in appearance of the so-called magneto-electric effects.  
It is worthwhile to mention that similar effects occur in axion like systems \cite{axion}. More precisely, Wilczec investigated the electrodynamics of topologically nontrivial pseudo-particles called \textit{axions} and demonstrated that variation in the axion field can produce some peculiar distributions of charge and current leading, in particular, to appearance of fractional electric charge on dyons and magneto-electric effects.  Thus, from the electrodynamics point of view, we can consider a TI as an axionic medium \cite{monopolo,witten,em} and, by analogy, add the following topological term to the conventional Maxwell action 
 \begin{equation}\label{axion}
 S_\theta=\frac{\alpha}{2\pi}\int d^3xdt\left({\bf E}\cdot{\bf B}\right)\theta(\textbf{r}),
\end{equation} 
 where $ \alpha$ is the fine structure constant, and $\theta $ is a phenomenological model parameter that characterizes the topological phase of TI. Recall that in High Energy Physics $\theta $ is known as an effective action term in the axion electrodynamics \cite{axion}.  With the introduction of the above axionic term, we arrive at the  constitutive relations describing the macroscopic properties of the TI, namely:

\begin{equation}\label{relatconst1}
\textbf{D}=\epsilon_0\epsilon(\textbf{r})\textbf{E}-\frac{\alpha}{2\pi}\theta(\textbf{r})\textbf{B},
\end{equation}

\begin{equation}\label{relatconst2}
\textbf{H}=\frac{1}{\mu_0\mu(\textbf{r})}\textbf{B}+\frac{\alpha}{2\pi}\theta(\textbf{r})\textbf{E},
\end{equation}
The modified constitutive relations  (\ref{relatconst1}) and (\ref{relatconst2}) are responsible for observation of \textit{magneto-electric effect} in $Hg Te$ in  $3D$ TI\cite{nature}.
To model a periodic photonic crystal, we assume that the above-introduced parameters $\epsilon$ (dielectric constant), $\mu$ (magnetic constant) and $\theta$ (topological number or axion coupling) are periodic with a spacial period equal to $d=a+b$, that is:

 \begin{equation}
   \left(
     \begin{array}{c}
        \epsilon(\textbf{r})\\
        \mu(\textbf{r})\\
        \theta(\textbf{r})
     \end{array}
    \right) =   
   \left(
     \begin{array}{c}
        \epsilon(z)\\
        \mu(z)\\
        \theta(z)
     \end{array}
    \right) =    
 \left(
     \begin{array}{c}
        \epsilon(z+a+b)\\
        \mu(z+a+b)\\
        \theta(z+a+b)
     \end{array}
    \right)   
 \end{equation}
 Fig.1 illustrates the proposed periodic photonic crystal based on topological insulators with a succession of repeating layers with different values of $\epsilon$, $\mu$, and $\theta$. 

\begin{figure}
\centerline{\includegraphics[scale=1.0]{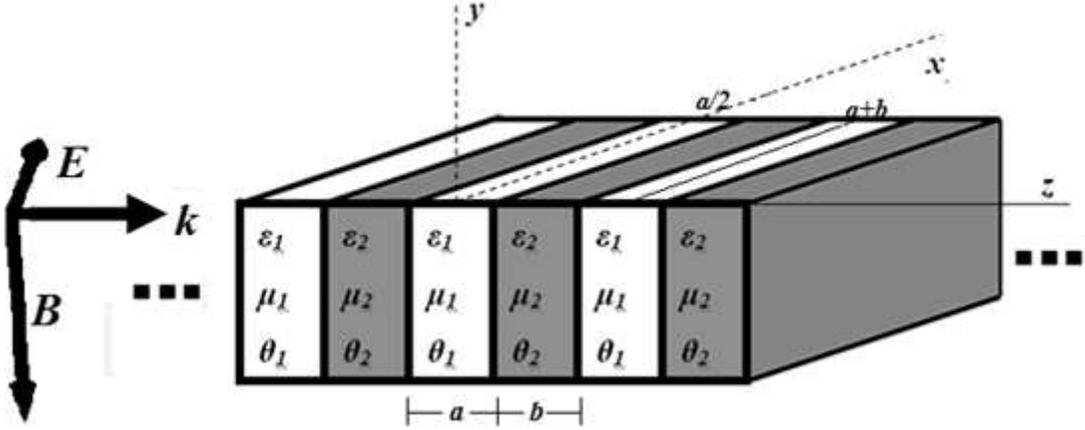} }
\caption{The proposed system: a photonic crystal composed of two layers with distinctive electric ($\epsilon$), magnetic ($\mu$) and topological ($\theta$)  properties.} 
\end{figure}

\section{Results and Discussion}

For the normal incidence of plane waves (with the wave vector $\textbf{k}=k\hat{\textbf{z}}$),  from the source-free Maxwell equations

\begin{equation}
\nabla\cdot\textbf{D}=0,
\end{equation}

\begin{equation}
\nabla\cdot\textbf{B}=0,
\end{equation}

\begin{equation}
\nabla\times\textbf{E}=-\frac{\partial \textbf{B}}{\partial t},
\label{faraday}
\end{equation}

\begin{equation}
\nabla\times\textbf{H}=\frac{\partial\textbf{D}}{\partial t},
\end{equation} 

we obtain 

\begin{equation}
\textbf{E}(\textbf{r},t)=e^{-i\omega t}(E_x(z)\hat{\textbf{x}}+E_y(z)\hat{\textbf{y}}), \ \ \ \ \ \ \textbf{B}(\textbf{r},t)=e^{-i\omega t}(B_x(z)\hat{\textbf{x}}+B_y(z)\hat{\textbf{y}}),
\end{equation}
for the field configurations of the problem, where (in view of Eq. (\ref{faraday}))

\begin{equation}
B_x=\frac{i}{\omega}\frac{dE_y}{dz}, \ \ \ \ \ B_y=-\frac{i}{\omega}\frac{dE_x}{dz}.
\label{magnetic}
\end{equation} 
Here,   $\omega$ is the  frequency of the plane wave.

Notice that in the regions where $\theta$ is uniform and constant, the Maxwell equations are identical to the ones for the conventional dielectrics. 

Since the parameters that characterize the medium are periodic, the fields should also obey the Bloch theorem \cite{jono}:
\begin{equation}
\textbf{E}(z+a+b)=e^{ik(a+b)}\textbf{E}(z), \ \ \ \ \ \ \textbf{B}(z+a+b)=e^{ik(a+b)}\textbf{B}(z),
\label{Bloch}
\end{equation} 
allowing us to restrict the problem to the unity cell defined as the region $0\leq z\leq a+b$. It can be easily verified that the general solution for both fields in this region reads

\begin{eqnarray}\label{Ex}
E_x=\left\{
            \begin{array}{lll}
               a_x\cos q_1z+b_x\sin q_1z,& & 0 \leq z<\frac{a}{2}\\  
               c_x\cos q_2z+d_x\sin q_2z,& & \frac{a}{2}\leq z<\frac{a}{2}+b\\
               e_x\cos q_1z+f_x\sin q_1z,& & \frac{a}{2}+b\leq z<a+b
            \end{array} \right.
\end{eqnarray}

\begin{equation}
E_y=\left\{
            \begin{array}{lll}
               a_y\cos q_1z+b_y\sin q_1z,& & 0\leq z<\frac{a}{2}\\
               c_y\cos q_2z+d_y\sin q_2z,& & \frac{a}{2}\leq z<\frac{a}{2}+b\\
               e_y\cos q_1z+f_y\sin q_1z,& & \frac{a}{2}+b\leq z<a+b
            \end{array}\right.
\label{Ey}
\end{equation}

\begin{equation}
B_x=\left\{
            \begin{array}{lll}
               \frac{i}{\omega}q_1a_y\sin q_1z  -  \frac{i}{\omega}q_1b_y\cos q_1z,& & 0\leq z<\frac{a}{2}\\
               \frac{i}{\omega}q_2c_y\sin q_2z  -  \frac{i}{\omega}q_2d_y\cos q_2z,& & \frac{a}{2}\leq z<\frac{a}{2}+b\\
               \frac{i}{\omega}q_1e_y\sin q_1z  -  \frac{i}{\omega}q_1f_y\cos q_1z,& & \frac{a}{2}+b\leq z<a+b
            \end{array}\right.
\label{Bx}
\end{equation}

\begin{equation}
B_y=\left\{
            \begin{array}{lll}
               -\frac{i}{\omega}q_1a_x\sin q_1z  +  \frac{i}{\omega}q_1b_x\cos q_1z,& & 0\leq z<\frac{a}{2}\\
               -\frac{i}{\omega}q_2c_x\sin q_2z  +  \frac{i}{\omega}q_2d_x\cos q_2z,& & \frac{a}{2}\leq z<\frac{a}{2}+b\\
               -\frac{i}{\omega}q_1e_x\sin q_1z  +  \frac{i}{\omega}q_1f_x\cos q_1z,& & \frac{a}{2}+b\leq z<a+b
            \end{array}\right.
\label{By}
\end{equation}
where $q_i=\frac{\omega\sqrt{\epsilon_i\mu_i}}{c}$ with $c$ being the velocity of light.

After applying the interface boundary conditions at $z=\frac{a}{2}$ and $z=\frac{a}{2}+b$, we finally obtain   
a transcendental equation as a condition for existence of nontrivial self-consistent solutions of the above system, namely  
\begin{equation}
\cos k(a+b)=\cos q_1a\cos q_2b-\Gamma(\delta)\sin q_1a\sin q_2b,
\end{equation}
where 
\begin{equation}
\Gamma(\delta)=\frac{1}{2}\left(\frac{Z_1}{Z_2}+\frac{Z_2}{Z_1}+ \delta Z_1Z_2\right),
\end{equation}
with $Z_i=\sqrt{\frac{\mu_i}{\epsilon_i}}$ being the impedance of the medium, and $\delta =\frac{\alpha\mu_0^2}{4\pi^2}\gamma^2$ the topological insulator induced gap related parameter with $\gamma=\theta_2-\theta_1$.

Recalling that $q_i=\frac{\omega}{c}\sqrt{\epsilon_i\mu_i}$, it becomes clear that Eq.(19) describes an implicit dependence of the frequency $\omega$ on the wave vector $k$. For a particular case when $\epsilon_1=\epsilon_2=\mu_1=\mu_2=1$ and $b=a$, Eq.(19) can be resolved explicitly leading to the following expression

\begin{equation}
\omega(k)=\omega_0\arccos \left[\frac{\delta +2\cos(\frac{\pi k}{k_0})}{\delta +2}\right],
\end{equation}
where $\omega_0=\frac{c}{2a}$ and $k_0=\frac{\pi}{2a}$.

The band diagram $\omega(k)$ for photonic crystal based on toplogical insulator for different values of $\delta$ is shown in Fig.2.  Notice that for conventional (nontopological) insulator with $\delta=0$ there is no gap in the spectrum. The gap opens only for non-zero values of $\delta$ and its width increases with increasing of $\delta$. 

\begin{figure}
\centerline{\includegraphics[width=5cm]{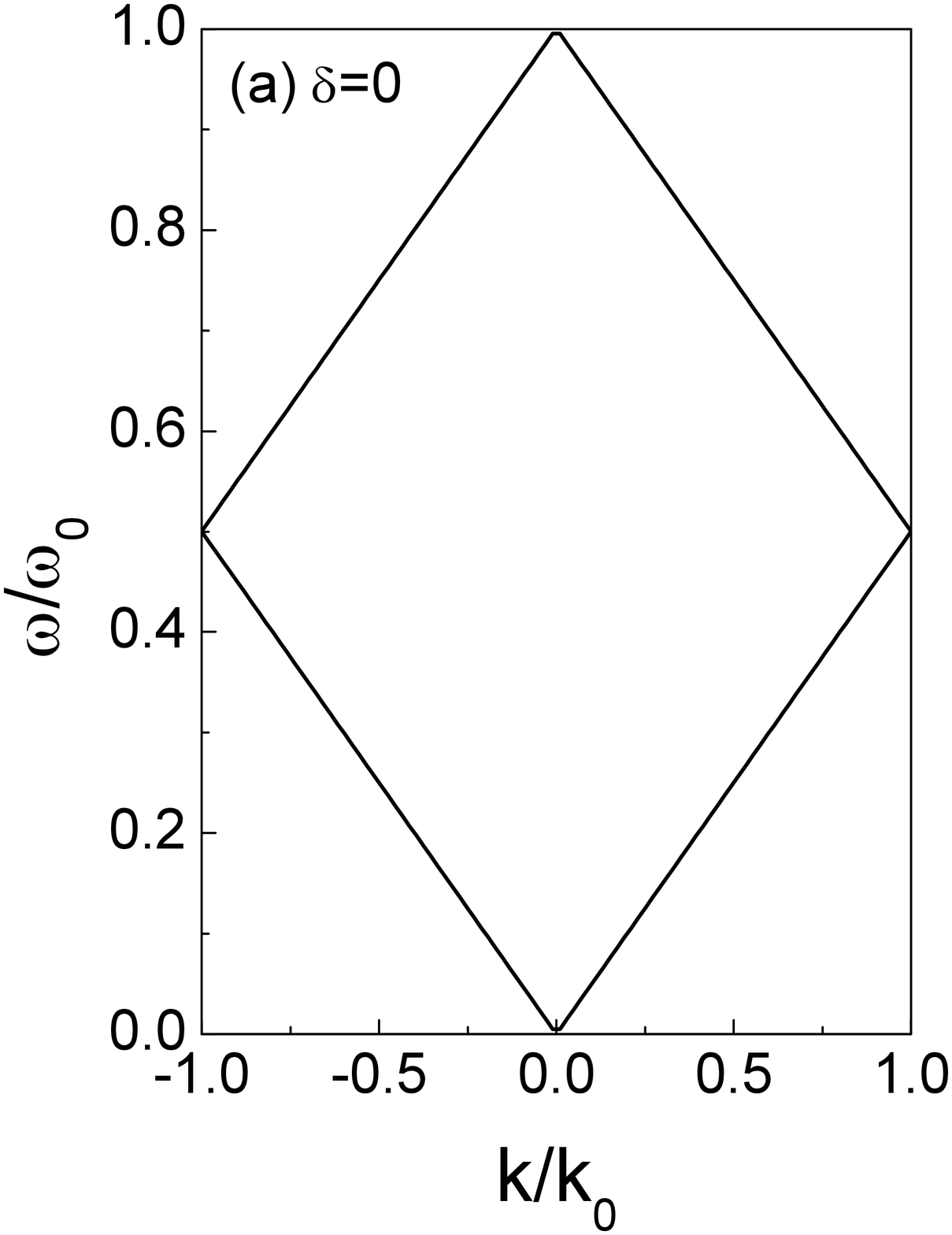} }\vspace{0.05cm}
\centerline{\includegraphics[width=5cm]{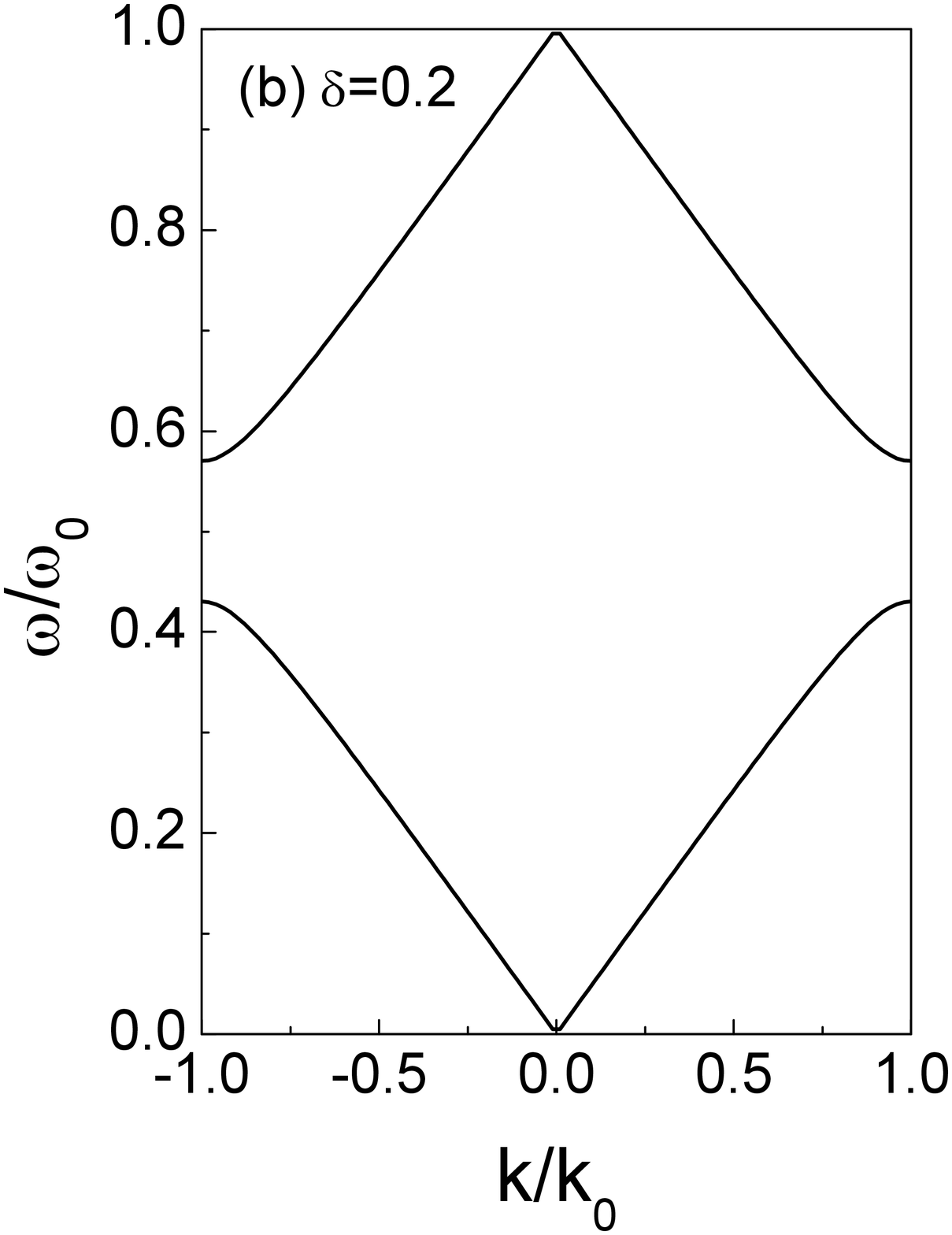} }\vspace{0.05cm}
\centerline{\includegraphics[width=5cm]{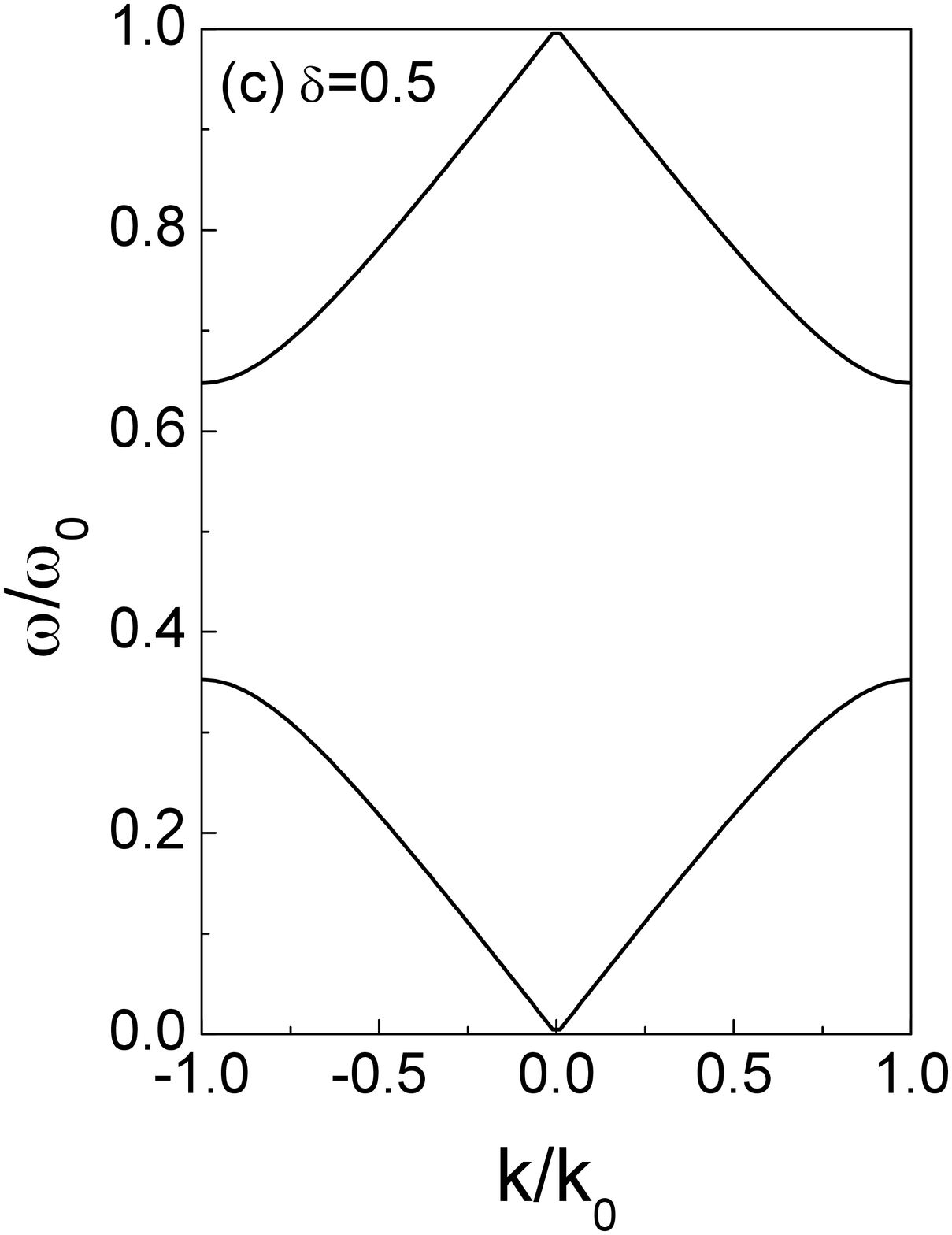} }\vspace{0.0cm}
\caption{The band diagram $\omega(k)$ of photonic crystal based on topological insulators for different values of the gap related parameter $\delta$ according to Eq.(21).}
\end{figure}

\begin{figure}
\centerline{\includegraphics[width=5.50cm]{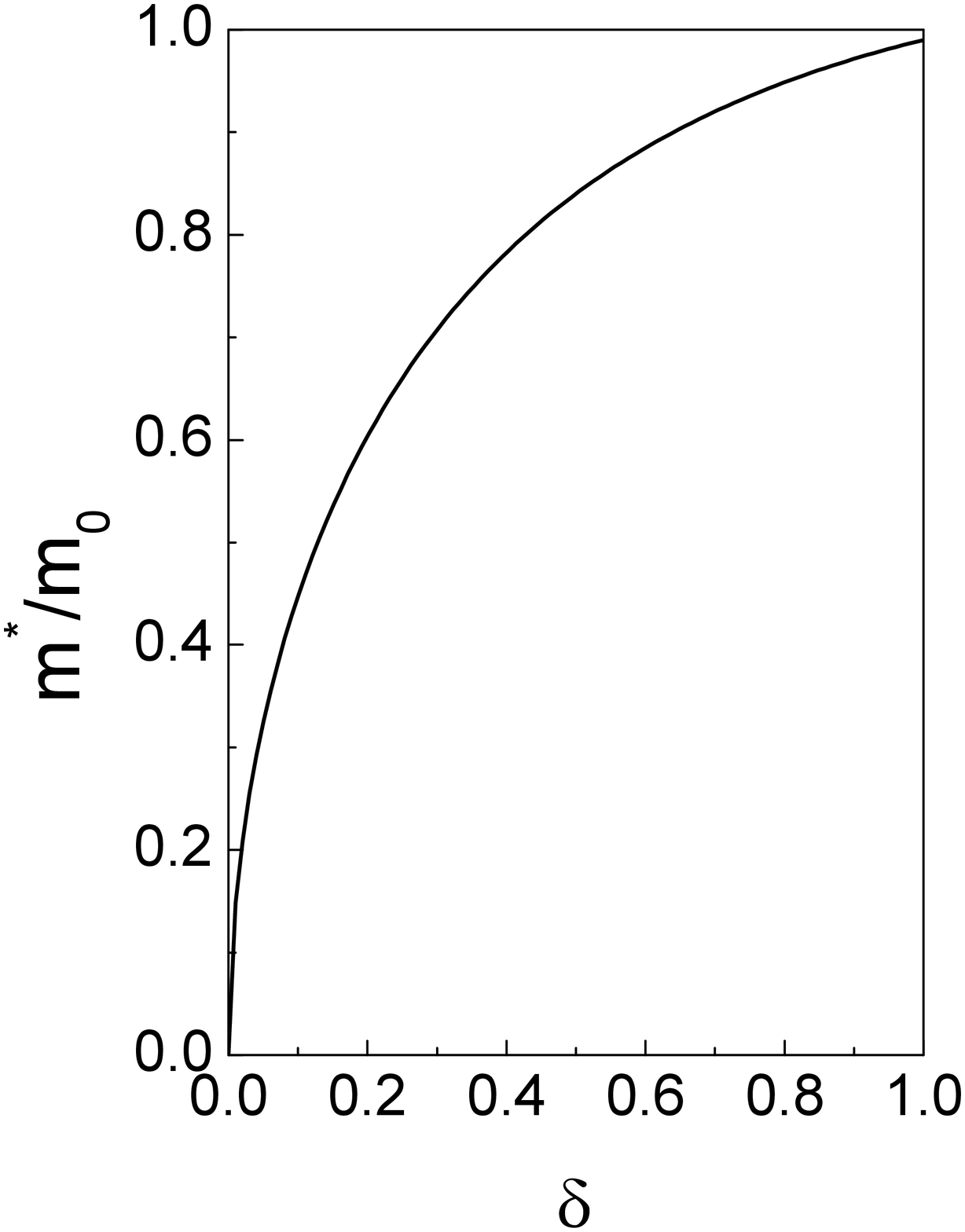} }\vspace{0.5cm}
\caption{The dependence of normalized effective photon mass on the gap related parameter $\delta$ according to Eq.(23).}
\end{figure}

Furthermore, using the above spectral law $\omega(k)$, we can introduce an effective photonic mass $m^{*}$, namely
\begin{equation}
\frac{1}{m^{*}}=\frac{1}{\hbar}\left(\frac{\partial ^2\omega}{\partial k^2}\right)_{k=k_0}.
\end{equation}
From Eq.(21) we obtain
\begin{equation}
m^{*}(\delta)=m_0\left(\frac{3\sqrt{\delta}}{\delta +2}\right),
\end{equation}
where $m_0=\frac{4\hbar}{3ac}$. 

The evolution of the normalized effective mass $m^{*}/m_0$ with the gap parameter $\delta$ is shown in Fig.3. It is also important to emphasize that (for a given value of $\delta$) the absolute value of the photon mass $m^{*}$ drastically depends on the period of the photonic crystal $d=2a$, ranging from light photons (with $m^{*}\simeq 10^{-36}kg$ for $d=1\mu m$) to heavy photons (with $m^{*}\simeq 10^{-33}kg$ for $d=1nm$). The latter estimate is typical for nontopological photonic crystals \cite{s8}. It would be interesting to experimentally check these  predictions using the existing PTI.

\section{Conclusion}

In summary, we studied propagation of light in a model system describing periodic photonic crystal comprised of topological insulators. By adding an axionic like term into conventional Maxwell equations, we obtained a closed system of linear equations on electric and magnetic fields. By resolving the resulting transcendental equation,  the existence of nontrivial band structure in the photonic topological insulator (PTI) was revealed. Namely, it was found that for nonzero values of the topological phase parameter $\gamma=\theta_2-\theta_1$ a finite gap  $\delta \propto\gamma^2$ opens in the spectrum which is equivalent to appearance of nonzero effective photon mass $m^{*}(\delta)$. The latter was shown to quite noticeably increase with $\delta$ and decrease with the period of the photonic crystal $d=a+b$.

{\bf Acknowledgements}

We thank Brazilian agencies CAPES, CNPQ  and FAPESQ for financial support.

\end{document}